\documentclass[12pt]{iopart}
\input xy
\xyoption{all}

\usepackage[pdftex]{graphicx}

\renewcommand{\S}{\mathcal{S}}

\begin{document}
\title[Geometry of quantum evolution]{Geometry of quantum evolution for mixed quantum states}
\author{Ole Andersson and Hoshang Heydari}
\address{Department of Physics, Stockholm  University, SE-106 91 Stockholm, Sweden}
\ead{olehandersson@gmail.com}

\begin{abstract}
The geometric formulation of quantum mechanics is a very interesting field of research which has many applications in the emerging field of quantum computation and quantum information, such as schemes for optimal quantum computers. In this work we discuss a geometric formulation of mixed quantum states represented by density operators. Our formulation is based on  principal fiber bundles and  purifications of quantum states.  In our construction, the Riemannian metric and symplectic form on the total space are induced from the real and imaginary parts of the Hilbert-Schmidt Hermitian inner product, and we define a mechanical connection in terms of a locked inertia tensor and moment map. We also discuss some applications of our geometric framework.
\end{abstract}

\section{Introduction}
Ever since the advent of general relativity, scientists have been looking for geometrical principles underlying physical laws.
Nowadays it is well known that geometry
affects the physics on all length scales, and physical theory building consists to a large extent of geometrical considerations.
This paper concerns geometric quantum mechanics, a branch of quantum physics that has received much attention lately -- largely due to the crucial role geometry plays in quantum information and quantum computing.
Here we equip the phase spaces
for unitarily evolving finite level quantum systems with natural Riemannian and symplectic structures, and
establish remarkable but fundamental relations between these and quantum theory.
Important previous works on the geometrical formulation of quantum mechanics that should be mentioned in this context are the book \cite{Bengtsson_etal2008} by Bengtsson and \.{Z}yczkowski and the papers \cite{Grabowski_etal2005} by J. Grabowski, M. K\'us and G. Marmo and \cite{Ashtekar_etal1998} by Ashtekar and Schilling.

A quantum system prepared in a pure state is usually modeled on a projective Hilbert space, and if the system is closed its state will evolve unitarily in this space.
The state of an experimentally prepared quantum system  generally
exhibits classical uncertainty, and is most appropriately described as a probabilistic mixture of pure states. It is common to represent mixed states by density operators, and many metrics on spaces of density operators have been developed to capture various physical, mathematical, or information theoretical aspects of quantum mechanics \cite{Bengtsson_etal2008, Nielsen_etal2010}.

In this paper we discuss a  geometrical framework for general quantum states represented by density operators on finite dimensional quantum systems. We show that our geometrical framework enable us to establish very important relations between abstract geometrical structures and general quantum systems which gives  breakthrough insight in our understanding of the foundations of quantum mechanics and quantum information with many applications. In section \ref{GfW} we give an introduction to our geometric framework for mixed quantum states, and in section \ref{applications} we discuss some recent applications  of the  framework such as  operational geometric phases \cite{GP}, a geometric uncertainty relation \cite{GUR}, and  a  dynamic distance measure \cite{DD}. This paper is based on \cite{GP,GUR,DD, ED,GS}.

\section{Geometry of orbits of isospectral density operators}\label{GfW}
In this paper we will only be interested in  finite dimensional quantum systems that evolve unitarily.
They will be modeled on a Hilbert space $\mathcal{H}$ of unspecified dimension $n$, and their states will be represented by density operators.
Recall that a density operator is a Hermitian, nonnegative operator with unit trace.
We write $\mathcal{D}(\mathcal{H})$ for the space of density operators on $\mathcal{H}$.

\subsection{Riemannian structure on orbits of density operators}
A density operator whose evolution is governed by a von Neumann equation remains in a single orbit of the left conjugation action of the unitary group of $\mathcal{H}$ on $\mathcal{D}(\mathcal{H})$. The orbits of this action are in one-to-one correspondence with the possible spectra for density operators on $\mathcal{H}$, where by the \emph{spectrum} of a density operator of rank $k$ we mean the decreasing sequence
\begin{equation}
\sigma=(p_1,p_2,\dots,p_k)
\label{spectrum}
\end{equation}
of its, not necessarily distinct, \emph{positive} eigenvalues.
Throughout this paper we fix $\sigma$, and write $\mathcal{D}(\sigma)$ for the corresponding orbit.

To furnish $\mathcal{D}(\sigma)$ with a geometry let $\mathcal{L}(\mathbf{C}^k,\mathcal{H})$ be the space of linear maps from $\mathbf{C}^k$ to $\mathcal{H}$, $P(\sigma)$ be
the diagonal $k\times k$ matrix
that has $\sigma$ as its diagonal, set
\begin{equation*}
\mathcal{S}(\sigma)=\{\Psi\in\mathcal{L}(\mathbf{C}^k,\mathcal{H}):\Psi^\dagger \Psi=P(\sigma)\},
\end{equation*}
and define
\begin{equation*}
\pi:\mathcal{S}(\sigma)\to\mathcal{D}(\sigma),\quad \Psi\mapsto\Psi\Psi^\dagger.
\label{bundle}
\end{equation*}
Then $\pi$ is a principal fiber bundle with right acting gauge group
\begin{equation*}
\mathcal{U}(\sigma)
=\{U\in\mathcal{U}(k):UP(\sigma)=P(\sigma)U\},
\end{equation*}
whose Lie algebra is
\begin{equation*}
\mathbf{u}(\sigma)
=\{\xi\in\mathbf{u}(k):\xi P(\sigma)=P(\sigma)\xi\}.
\end{equation*}
We equip $\mathcal{L}(\mathbf{C}^k,\mathcal{H})$ with the Hilbert-Schmidt Hermitian product, and $\S(\sigma)$ with the Riemannian metric $G$ and the symplectic form $\Omega$ given by $2\hbar$ times the real and imaginary parts, respectively, of this product:
\begin{equation}
G(X,Y)=\hbar\mathrm{Tr}(X^\dagger Y+Y^\dagger X), 
\qquad
\Omega(X,Y)=-i\hbar\mathrm{Tr}(X^\dagger Y-Y^\dagger X). 
\end{equation}
We also equip $\mathcal{D}(\sigma)$ with the unique metric $g$ that makes $\pi$ a Riemannian submersion.

\subsection{Mechanical connection}
The \emph{vertical} and \emph{horizontal bundles} over $\mathcal{S}(\sigma)$ are the subbundles
$\mathrm{V}\mathcal{S}(\sigma)=\mathrm{Ker} \pi_*$ and $\mathrm{H}\mathcal{S}(\sigma)=\mathrm{V}\mathcal{S}(\sigma)^\bot$
of the tangent bundle of $\mathcal{S}(\sigma)$.
Here $\pi_*$ is the differential of $\pi$ and $^\bot$ denotes orthogonal complement with respect to $G$.
Vectors in $\mathrm{V}\mathcal{S}(\sigma)$ and $\mathrm{H}\mathcal{S}(\sigma)$
are called vertical and horizontal, respectively,
and a curve in $\mathcal{S}(\sigma)$ is called horizontal if its velocity vectors are horizontal.
Recall that for every curve $\rho$ in $\mathcal{D}(\sigma)$ and every $\Psi_0$ in the fiber over $\rho(0)$ there is a unique horizontal lift of $\rho$ to $\mathcal{S}(\sigma)$ that extends from $\Psi_0$.
This lift and $\rho$ have the same lengths because $\pi$ is a Riemannian submersion.

The infinitesimal generators of the gauge group action yield canonical isomorphisms between $\mathbf{u}(\sigma)$ and the fibers in $\mathrm{V}\mathcal{S}(\sigma)$:
\begin{equation}\label{eq:inf gen}
\mathbf{u}(\sigma)\ni\xi\mapsto \Psi\xi\in\mathrm{V}_\Psi\mathcal{S}(\sigma).
\end{equation}
Furthermore, $\mathrm{H}\mathcal{S}(\sigma)$ is the kernel bundle of the gauge invariant \emph{mechanical connection form}
$\mathcal{A}_{\Psi}=\mathcal{I}_{\Psi}^{-1}J_{\Psi}$,
where $\mathcal{I}_{\Psi}:\mathbf{u}(\sigma)\to \mathbf{u}(\sigma)^*$ and $J_{\Psi}:\mathrm{T}_{\Psi}{\mathcal{S}(\sigma)}\to \mathbf{u}(\sigma)^*$ are the \emph{moment of inertia} and \emph{moment map}, respectively,
\begin{equation*}
\mathcal{I}_{\Psi}\xi\cdot \eta=G(\Psi\xi,\Psi\eta),\qquad
J_{\Psi}(X)\cdot\xi=G(X,\Psi\xi).
\end{equation*}
The moment of inertia is of \emph{constant bi-invariant type} since it is an adjoint-invariant form on $\mathbf{u}(\sigma)$ which is independent of $\Psi$ in $\mathcal{S}(\sigma)$. To be exact,
\begin{equation}\label{eq1}
\mathcal{I}_{\Psi}\xi\cdot \eta=\frac{1}{2}\mathrm{Tr}\left(\left(\xi^\dagger \eta+\eta^\dagger \xi\right)P(\sigma)\right).
\end{equation}
Using equation (\ref{eq1}) we can derive an explicit formula for the connection form.
Indeed, if $m_1, m_2, \dots , m_l$ are the multiplicities of the different eigenvalues in $\sigma$, with $m_1$ being the multiplicity of the greatest eigenvalue, $m_2$ the multiplicity of the second greatest eigenvalue, etc., and if for $j=1,2,\dots,l$,
\begin{equation*}
E_j=\mathrm{diag}(0_{m_1},\dots,0_{m_{j-1}},1_{m_j},0_{m_{j+1}},\dots,0_{m_l}),
\end{equation*}
then
\begin{eqnarray}
\mathcal{I}_\Psi(\sum_jE_j\Psi^\dagger XE_jP(\sigma)^{-1})\cdot\xi
&=&
\frac{1}{2}\mathrm{Tr}(\sum_jE_jX^\dagger\Psi E_j\xi-\xi E_j\Psi^\dagger XE_j)\\\nonumber&=&
\frac{1}{2}\mathrm{Tr}(X^\dagger\Psi\xi-\xi\Psi^\dagger X)\\\nonumber&=&
J_\Psi(X)\cdot\xi
\end{eqnarray}
for every $X$ in $\mathrm{T}_\Psi\mathcal{S}(\sigma)$ and every $\xi$ in $\mathbf{u}(\sigma)$.
Thus,
\begin{equation*}\label{eq:explicit}
\mathcal{A}_\Psi(X)=\sum_jE_j\Psi^\dagger XE_jP(\sigma)^{-1}.
\end{equation*}
Observe that the orthogonal projection of $\mathrm{T}_\Psi\mathcal{S}(\sigma)$ onto $\mathrm{V}_\Psi\mathcal{S}(\sigma)$
is given by the connection form followed by the infinitesimal generator given by equation (\ref{eq:inf gen}). Therefore, the \emph{vertical} and \emph{horizontal projections} of $X$ in $\mathrm{T}_\Psi\mathcal{S}(\sigma)$ are $X^\bot=\Psi\mathcal{A}_\Psi(X)$ and $X^{||}=X-\Psi\mathcal{A}_\Psi(X)$, respectively.

\section{Applications of geometric framework for mixed quantum states}\label{applications}
We have introduced a geometrical framework for mixed quantum states represented by density operators which has so far resulted in an operational geometric phase and higher order
geometric phases, a geometric uncertainty relation, a dynamic distance measure, and an energy estimate and a classification of optimal Hamiltonians for mixed quantum states. In this section we will briefly discuss these applications of our framework.

\subsection {Operational geometric phases for non-degenerate and degenerate mixed quantum states}
Geometric phases are very important tools both in classical and quantum physics. 
Uhlmann \cite{Uhlmann_1976,Uhlmann1986, Uhlmann1989, Uhlmann1991} was among the first to 
develop a theory for geometric phase for parallel transported mixed states.
The theory is based on the concept of purification.
Another approach to geometric phase for parallel transported non-degenerate mixed states, based on quantum interferometry, was proposed by Sj\"{o}qvist \emph{et al.} \cite{Sjoqvist_etal2000}. This phase has been verified in several experiments, and according to Slater \cite{Slater2002} it generally yields different outcomes than that of Uhlmann.

Recently,  we have introduced an operational geometric phase \cite{GP} for mixed quantum states based on spectral weighted traces of holonomies, and we have shown that it generalizes the interferometric definition of Sj\"{o}qvist \emph{et al.} The operational geometric phase is a direct application of the Riemannian structure of our geometric framework. We also introduce higher order geometric phases for mixed quantum states. Our operational geometric phase applies to general unitary evolutions of  nondegenerate and also degenerate mixed states.
The operational geometric phase is defined by
\begin{equation}
\gamma_{g}(\rho)=
\arg \mathrm{Tr}(\Psi^{\dagger}_{0}\Pi[\rho]\Psi_{0})\nonumber=
\arg \mathrm{Tr}(\Psi^{\dagger}_{\|}(0)\Psi_{||}(\tau)),
\end{equation}
where
$\Pi[\rho]\Psi_{0}=\Psi_{\|}(\tau)$
and $\Psi_{\|}$ is the horizontal lift of $\rho$ extending from $\Psi_{0}$:
\begin{equation}\label{horiz}
\Psi_{||}(t)=\Psi(t)\exp_{+}\left(-\int_{0}^{t}\mathcal{A}_\Psi(\dot \Psi)\, dt\right).
\end{equation}
Here $\exp_{+}$ is the positive time-ordered exponential. For more details about the construction of the operational geometric phase and higher order geometric phases we refer the reader to our recent paper \cite{GP}.

\subsection{Dynamic distance measure}
Distance measures are very important tools in quantum information processing.
Recently we have proposed a new distance measure for mixed quantum states that we call the dynamic distance measure.
The dynamic distance measure is defined in terms of a measurable quantity, which make it very suitable for applications.

Let $\rho_0$ and $\rho_1$ be isospectral density operators and consider a von Neumann equation
\begin{equation}\label{von}
i\dot\rho=[H,\rho],\qquad \rho(t_0)=\rho_0,\quad\rho(t_1)=\rho_1.
\end{equation}
We define
\begin{equation}
\mathcal{D}(H,\rho_0,\rho_1)=\int_{t_0}^{t_1}
\sqrt{\mathrm{Tr}(H^2\rho)-\mathrm{Tr}(H\rho)^2}\mathrm{dt}
\end{equation}
provided $H$ is such that a solution curve $\rho$ to (\ref{von}) exists, and we define
 the dynamic distance between $\rho_0$ and $\rho_1$ by
\begin{equation}\label{distance}
\mathrm{Dist}(\rho_0,\rho_1)=\inf_{H}\mathcal{D}(H,\rho_0,\rho_1),
\end{equation}
where the infimum is taken over all Hamiltonian $H$ for which the boundary value problem (\ref{von}) is solvable.
In \cite{DD} we have shown that (\ref{distance}) is a proper distance measure between isospectral density operators.
In fact, it is the distance function associated with the metric $g$.
We have also compared our dynamic distance measure with the well-known Bures distance \cite{Bures1969, Uhlmann1992, Dittmann1993,Dittmann1999}, and it turns out that the dynamic measure is bounded from below, but is in general not equal to, the Bures distance. The reason is that Uhlmann's definition of parallel transport is different from ours.

\section{Conclusion}
Recently,  we have used the framework presented in section \ref{GfW} to derive a geometric uncertainty relation for observables acting on mixed quantum states. For pure states the uncertainty relation reduces to the geometric interpretation of the Robertson-Schr\"{o}dinger uncertainty relation by Ashtekar and Schilling \cite{Ashtekar_etal1998}. But in general the two relations are not equivalent. This is due to the multiple dimensions of the gauge group for general mixed states. More information about our result, especially a comparison with the Robertson-Schr\"{o}dinger uncertainty relation, can be found in \cite{GUR}.

We have introduced a geometrical framework for general quantum states represented by density operators on finite dimensional quantum systems in mixed states that evolve unitarily. Our geometrical framework enable us to establish relation between geometrical structures and general quantum systems. This correspondence between geometry and quantum physics gives  new insight in our understanding of the foundations of quantum mechanics and quantum information with many applications. We have shown that our geometric framework has already resulted in a new operational geometric phase and higher order geometric phases and a new dynamic distance measure. There are other applications of our framework that worth mentioning such as  quantum speed limit and optimal quantum control for mixed quantum states. Unfortunately, there is no space left here to discuss these issues. Interested reader may see our paper \cite{ED} for further information and a detailed discussion of the subject.
We believe our geometric framework could results in many other interesting applications in the field of quantum dynamics,  quantum information, quantum computations, quantum control, and quantum optics.

\ack{This work was supported by the Swedish Research Council (VR).}

\end{document}